\documentstyle[prl,aps,psfig,preprint]{revtex}
\begin{document}


\title{Direct Observation of Lateral Coupling Between Self-Assembled 
Quantum Dots}

\author{H. D. Robinson$^1$, B. B. Goldberg$^1$, J. L. Merz$^2$}
\address{$^1$Dept. of Physics, Boston University, Boston MA 02215}
\address{$^2$Dept. of Electrical Engineering, 
Notre Dame University, South Bend IN 46556}
\date{\today}

\maketitle

\begin{abstract}
Scattering of carriers between spatially separated zero dimensional 
states has been observed in a 
system of self-assembled In$_{0.55}$Al$_{0.45}$As quantum dots. We believe
the interdot tunneling is mediated by localized states 
below the barrier band edge. The experiment was
performed by taking photoluminescence excitation spectra at 4.2~K using a 
near-field scanning optical microscope. Surprisingly, the
excitation spectrum from individual dots does not quench to zero at any 
energy, even when the 
energy of the exciting light is tuned below the barrier band edge.
On top of this continuum, narrow resonances are observed in the emission lines
of individual dots. These resonances tend to occur simultaneously in several 
emission lines, originating from different quantum dots, evincing 
interdot scattering.
\end{abstract}

\pacs{73.20.Dx,78.66.Fd,78.55.Cr}

Over the last few years, intense efforts have yielded a very detailed 
understanding of the physics of coupling between the electronic states of 
zero-dimensional (0D) quantum dots (QD)
\cite{klimeck-94-1,vaart-95-1,blick-96-1,schmidt-97-1,oosterkamp-98-1}. 
These processes
are much less studied in 0D systems where the states are excitonic rather than 
electronic, and only one study\cite{schedelbeck-97-1}, on a sample of cleaved 
edge overgrowth QDs, has reported a direct observation of coupling between 
individual QDs. In this Letter, we present the first such observation for the
important case of self-assembled quantum dots (SADs)\cite{petroff-94-1,moison-94-1,marzin-94-1,grundmann-95-1}. 

SADs form when an epitaxial layer of a semiconductor is grown on substrate with
which it has a large lattice mismatch (Stranski-Krastanov growth mode) 
\cite{stranski-39-1,goldstein-85-1}. When grown beyond
a critical thickness, the epitaxial layer spontaneously relieves the
strain by forming 3D islands of relatively uniform size. The result is
a layer of homogeneous, randomly distributed, high quality
quantum dots connected by a thinner quantum well known as the wetting
layer (WL). The growth method allows no control of dot positions, and the
dot density is typically very high, $\sim 10^{10}$~cm$^{-2}$. Therefore,
previous optical experiments have been largely limited to study 
low dot density samples in order to overcome the inevitable inhomogeneous
broadening caused by studying a large ensemble of dots. In such systems,
coupling between individual dots is unobservable. Using a near-field 
scanning optical microscope (NSOM)\cite{betzig-92-1,paesler-96-1} operating at 
4.2~K, we can study a small ensemble (10-25) of dots in samples where the dot 
density is sufficiently high for interdot coupling to occur.

The sample used consists of a In$_{0.55}$Al$_{0.45}$As quantum
well, containing the dots, embedded in 
Al$_{0.35}$Ga$_{0.65}$As grown on a GaAs substrate, and has been studied
extensively\cite{fafard-95-1,leon-95-1,leon-95-2,wang-96-1}. 
The dot density is known to be 
$2\times 10^{10}$~cm$^{-2}$, the average lateral dot radius ($R$) is 9~nm,
and the ground state and the first excited state separation 
($\hbar\omega_{01}$) is approximately 40~meV.

Photoluminescence excitation (PLE) measurements were carried out in the
near-field, illuminating the sample with the NSOM tip, and collecting the
emission with conventional optics in the far-field. This arrangement produces
the largest signal, but has in general a limited spatial resolution due to
diffusion of carriers away from the tip. Here, however, the data has
been gathered with the excitation energy below the wetting layer band edge, 
where such diffusion is very small. The PLE excitation and detection regions
are shown schematically in Fig.~\ref{fig:spectrum} on a 
photoluminescence spectrum of dots and the wetting layer exciton.

Fig.~\ref{fig:tail} shows the PLE spectra for two individual QD emission lines 
(sharp lines between 1840~meV and 1940~meV in Fig.~\ref{fig:spectrum}) as the 
laser is tuned through the WL exciton. 
The spectra show two broad peaks which can be identified with the heavy-hole
and light-hole excitons respectively. The most striking feature, however, is
the fact that both lines remain non-zero even when the excitation energy is
tuned to energies below the WL band-edge. This tail is present in the majority
of emission lines that we have observed. As the excitation energy is decreased
further, the signal intensity slowly decreases, typically by 10\% to 
25\% every 10 meV. 

When the laser is tuned more than 
20 meV below the WL band-edge, another striking feature appears in the
PLE spectra in the form of sharp resonances that begin to appear on top of 
the continuous background signal.
Fig.~\ref{fig:ress}a plots emission intensity as a function of both excitation
(ordinate) and detection (abscissa) showing the QD emission lines (vertical 
dark lines) with a number of resonances (darker spots on the lines). 
The FWHM of typically 0.2--0.4~meV means the resonances 
originate from discrete QD states. While it is natural to assume these 
are the excited states of the individual quantum dots, one
notices that resonances occur in several emission lines at the {\it exact}
same excitation energy. This is illustrated in Fig.~\ref{fig:ress}b by
spectral line cuts at energies indicated by arrows in Fig.~\ref{fig:ress}a.

We attribute the multiline resonances to a mechanism that allows
excitons to scatter between discrete states in separate but nearby dots,  
giving rise to emission from the ground state of different dots at the same 
excitation energy. In order to arrive at this conclusion, we
must first exclude several other possible explanations.

In several recent papers\cite{landin-98-1,dekel-98-1,chavez-perez-98-1}, 
emission from multiexciton states has
been observed at high pump intensities. These states manifest themselves in
the PL spectrum as several extra lines a few meV above and below the single
exciton line. In NSOM experiments, the optical power density impacting on the
very small area immediately underneath the tip can typically reach the
$10^3$--$10^4$~W/cm$^2$ range, making it quite reasonable to consider such
non-linear effects. 

In order to identify possible non-linearities, PL spectra were taken as a 
function of optical power. Two series of such spectra varying power over two 
orders of magnitude are shown in Fig.~\ref{fig:powbWL}b and c. Note that 
intensities have been divided by incident power, so that linear power 
dependence appears as a constant shade of gray. Plainly, almost all observed 
emission  lines scale linearly with power, both on and off resonance, except 
for a slight tendency to saturation at higher powers.
When exciting above the WL band edge, non-linearities similar to
what has been reported elsewhere\cite{chavez-perez-98-1} are observed in some 
emission lines. These lines
disappear when tuning below the band-edge. We can therefore safely conclude
that the various lines observed to resonate simultaneously are not due to
different multiexciton states of the same dot. Also, the fact that the 
power dependence is linear on resonances means that the excitation and
coupling processes themselves are linear, excluding for example two-photon
or state-filling effects.  

Previous work has shown there to be no perceptible phonon 
bottleneck\cite{bockelmann-90-1} in SADs,
leading to photoluminescence exclusively from the groundstate\cite{wang-94-1}.
If the power of the pump light is high enough to cause state filling of the
ground state, emission from higher lying states can occur\cite{raymond-96-1}. 
Since it is clear from Fig.~\ref{fig:powbWL} that no significant saturation 
and therefore no state filling effects are observed, we can
exclude the possibility of recombination from different excited states
of the same dot as an explanation for the multiline resonances. Moreover, the
separation between the multiresonance lines varies essentially randomly from
0.5~meV to over 25~meV. This is inconsistent both with the 40~meV excited state
separation in this sample and any level splitting due for example to deviation 
from cylindrical symmetry in the dots, which reaches at most a few meV.

Having excluded non-linear effects and excited states of the same dot, we 
conclude that each line in a set of multiresonant lines originate from a
separate quantum dot, and consequently there exists a mechanism 
for scattering excitons between dots. Indications of this type of behavior
has been reported in SAD systems with much higher dot 
density\cite{huffaker-98-1}, but this experiment shows the first direct 
evidence of this effect. 

In order to estimate the probability of direct dot-to-dot tunneling as a
mechanism for interdot scattering we have carried out a simple 
Monte-Carlo simulation of the spatial dot distribution
for the sample, assuming hard-wall potentials the size of the dots preventing 
overlap, but otherwise allowing equal probability for all dot separations.
This random dot distribution is consistent with experimental observations.
\cite{moison-94-1,leonard-94-1}
We find that 70\% of the time, the nearest neighbor edge-to-edge separation
is larger than 10~nm (86\% for 5~nm).

The tunneling time can be estimated in the small coupling limit to be
$\tau = {h\over 2\alpha}$, where $\alpha = \int \varphi_0({\bf r}) V_0({\bf r})
\varphi_1({\bf r})d{\bf r}$. By taking the potential 
$$V({\bf r}) = \left\{
\begin{array}{cl}
{1\over 2}m\omega_{01}^2(r^2-R^2) & r<R \\
0 & R>r \end{array}\right.$$
 and the wavefunction $\varphi({\bf r})$ to
be the solution to the harmonic oscillator for $r<R$ and the appropriate 
combination of Bessel functions $K_0$ and $K_2$ for $r>R$, we can estimate 
$\tau$ for tunneling between second excited states. We have used $m =$ 0.2, 
$R =$ 9~nm, and $\hbar\omega_{01} =$ 37~meV, which are close to the measured 
values, and yields a confinement of 130~meV, consistent with the measured 
distance between WL and QD emission. We find $\tau =$ 1~ns for a 10~nm dot 
separation, which is of the same order as the exciton recombination time.
This would mean that by our estimate about 30\% of the dots can be involved in 
interdot tunneling, which is consistent with the data. However, if the energy 
difference between initial and final states is large compared to their widths,
which is almost always the case, the tunneling must be assisted by emission or 
absorption of a LA phonon in order to conserve energy. This will make the 
process orders of magnitude slower, implying that direct tunneling between dots
is probably too weak to fully explain what we have observed.

The simplest way to explain the observations without invoking direct tunneling
would be to place localized states
inside the barrier between the dots. If the density of such states is high
enough, they can act as intermediate states for carriers moving between dots.
This would also explain why the signal remains finite when exciting below
the WL band edge, as the states form a ``tail'' in the density of states 
inside the bandgap, analogous to the situation in glasses. In a two dimensional
system, any type of disorder, e.g. interface roughness and alloy
fluctuations, will give rise to localized states. Such disorder is certainly
present, and if it is severe enough, states in the WL may localize on a 
length scale small enough for this explanation to apply. 

Experimental evidence supports this hypothesis. Firstly, 
In PL, the peak associated with the wetting layer exciton is unusually wide, 
and has a fair amount of structure in the near-field. An example of this is
shown in Fig.~\ref{fig:spectrum}. This could be taken as an indication of
the non-2D nature of the WL. Secondly, as can be seen in 
Fig.~\ref{fig:roughstates}, a close examination of PL spectra at energies
immediately below the WL exciton peak reveals a multitude of very weak, 
narrow lines, the brightest of which show strong saturation at high excitation
powers. This is a clear indication of the presence of localized states.    

Since the observed resonances are consistent with scattering from the
second excited state of the dots, no resonances being observed at higher
or lower energies, and since the peaks observed between WL and QD emissions
are so weak and easy to saturate, we can conclude that we are indeed observing 
scattering from one dot to another, rather than from one WL state into
several nearby dots. 

We finally note two possible consequences of our observations for the 
physics of SAD systems, both due to new carrier relaxation channels being
opened by interdot coupling. We have already noted that SADs display no phonon 
bottleneck, and several explanations for this have been suggested, including
multiphonon emission\cite{inoshita-92-1,ohnesorge-96-2,heitz-97-1} and 
scattering off of holes\cite{sosnowski-98-1} or deep-level
impurities\cite{sercel-95-1}.  It is possible that interdot scattering 
provides an additional explanation for this, which moreover sets SADs apart 
from related 0D systems where a phonon bottleneck has been 
observed\cite{hasen-97-1}. Another possible consequence
is that state-filling at high optical powers may be more difficult to observe.
Indeed, far-field experiments at much higher powers than used in the near-field
show no emission from higher excited states, and we 
tentatively suggest this as a method, not involving high spatial
resolution techniques, for testing the strength of lateral interdot coupling
in self-assembled quantum dot systems.

This work was supported by NSF Grant No. DMR-9701958. The authors gratefully
acknowledge P. M. Petroff, whose laboratory produced the sample used for
these studies. The authors also thank Pawel Hawrylak for helpful discussions.

\vspace{0.2in}

\bibliography{press}
\bibliographystyle{prsty}

\newpage

\begin{figure}
\caption{A typical photoluminescence spectrum in the near-field, where 
emission lines from individual quantum dots can be resolved. Detection and
excitation regions for collecting PLE data are schematically indicated. The 
feature around 1995~meV is the wetting layer exciton.
\label{fig:spectrum}}
\end{figure}

\begin{figure}
\caption{A comparison of the PLE of two quantum dot emission lines with the
PL of the wetting layer exciton. There are no vertical offsets of the graphs.
\label{fig:tail}}
\end{figure}

\begin{figure}
\caption{(a) Typical plot of intensity vs. laser- and detection-energy. 
Numerous resonances are visible. The arrows mark vertical line cuts shown in
 (b).
(b) Line cuts from a, offset vertically for clarity.
\label{fig:ress}}
\end{figure}

\begin{figure}
\caption{(a) PLE of strong multiline resonance. Arrows indicate excitation
energies for power dependence off resonance, shown in (b), and on resonance, 
shown in (c). \label{fig:powbWL}}
\end{figure}

\begin{figure}
\caption{PL with $\lambda_{exc}$ = 514.5~nm. The spectra have been scaled
inversely to power and offset slightly to enable mutual comparison. The 
heights of the WL exciton peaks at 1995~meV are about eight times the full 
vertical scale of the figure.
\label{fig:roughstates}}
\end{figure}

\newpage
\psfig{figure=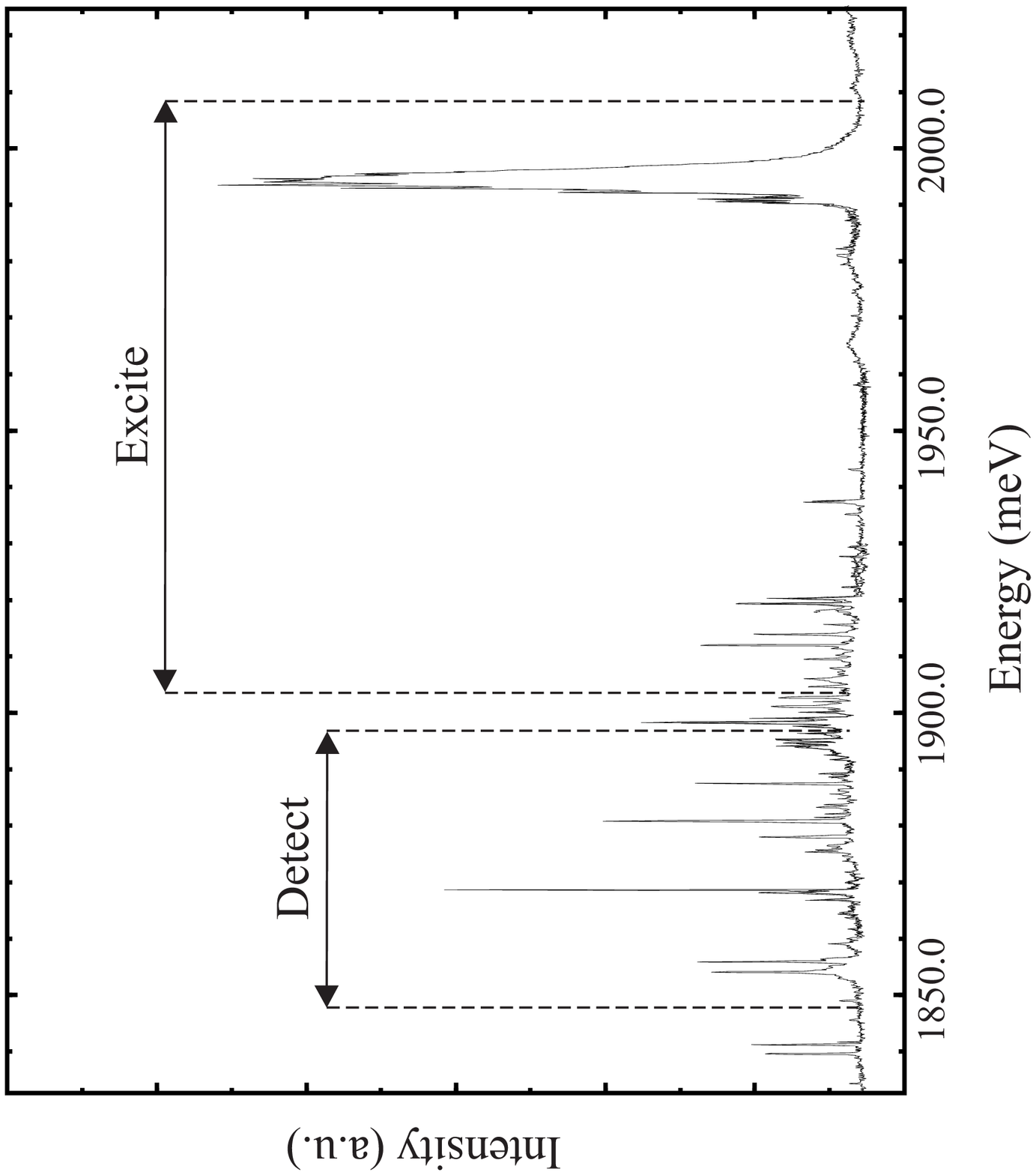,angle=90}
\begin{center}
  Figure \ref{fig:spectrum}
\end{center}

\newpage
\psfig{figure=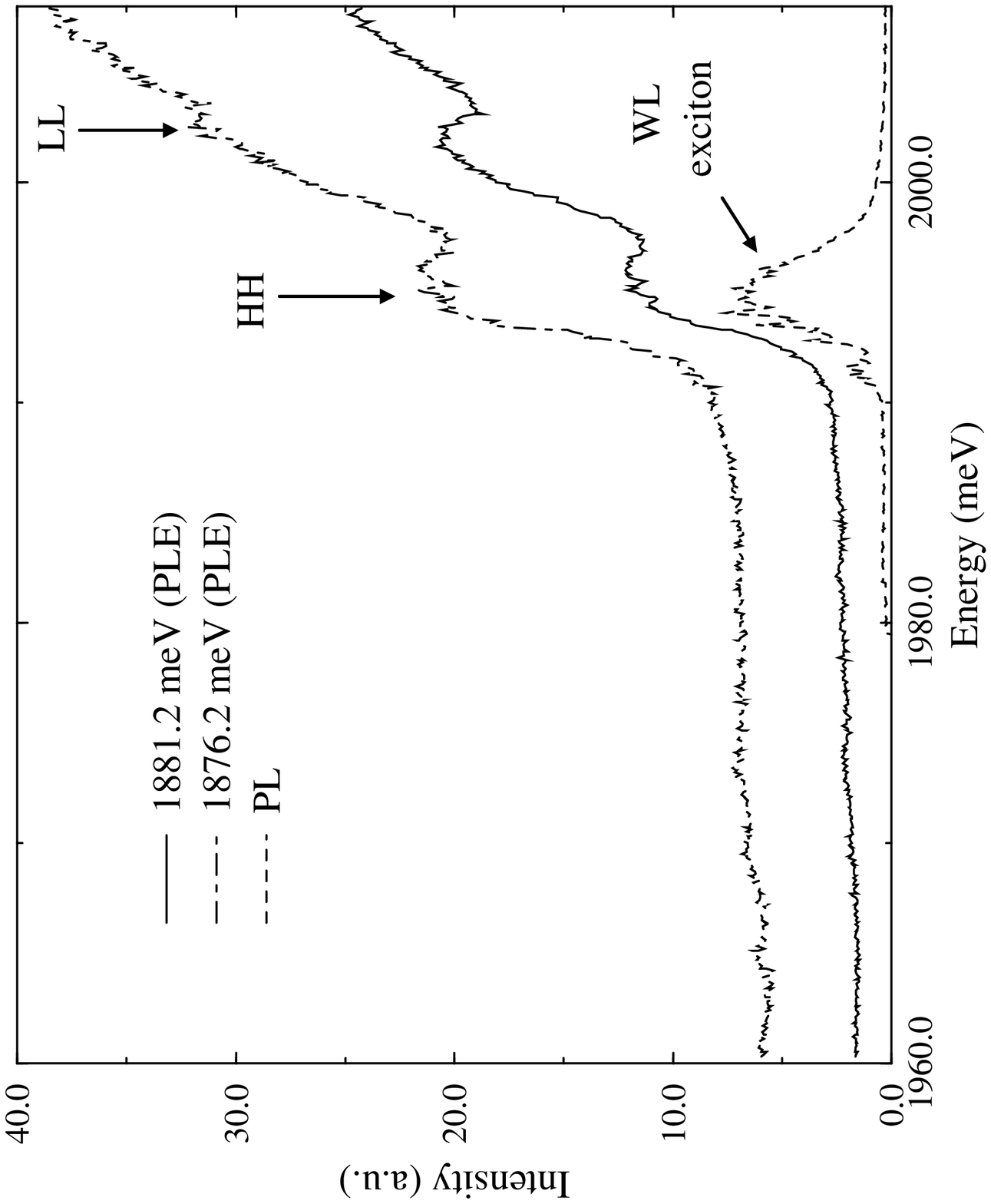}
\begin{center}
  Figure \ref{fig:tail}
\end{center}

\newpage
\psfig{figure=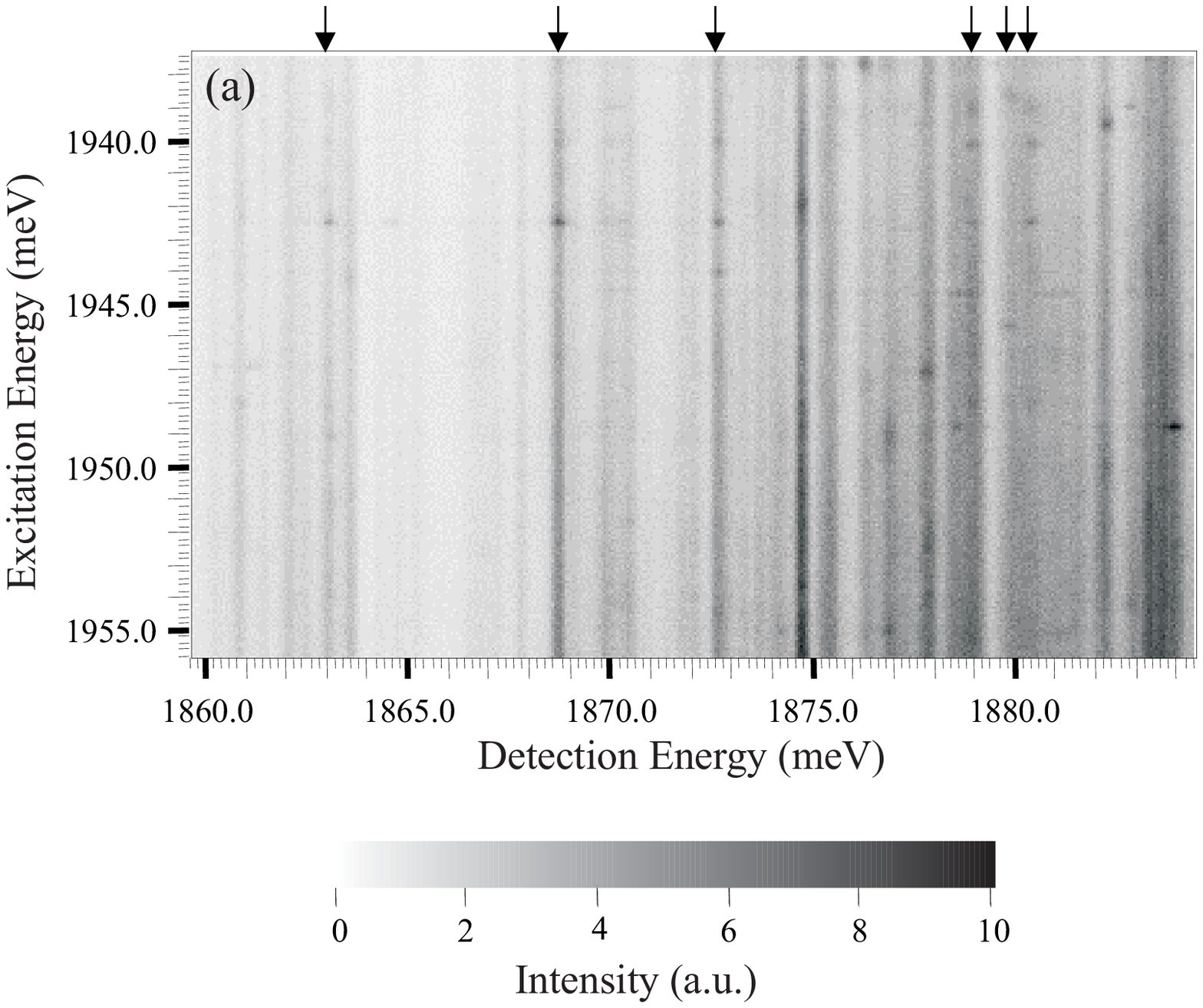}
\begin{center}
  Figure \ref{fig:ress}a
\end{center}

\newpage
\psfig{figure=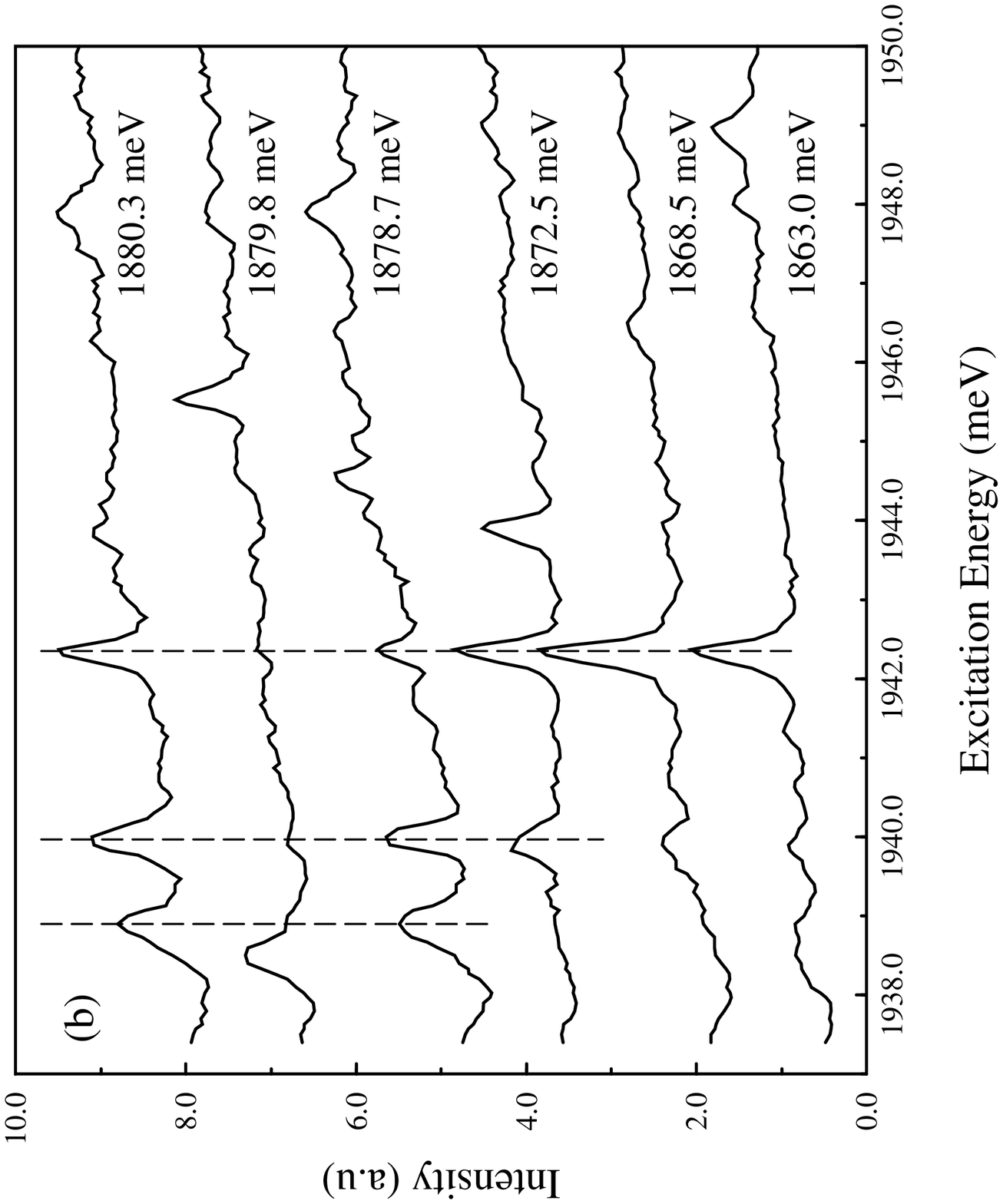}
\begin{center}
  Figure \ref{fig:ress}b
\end{center}

\newpage
\psfig{figure=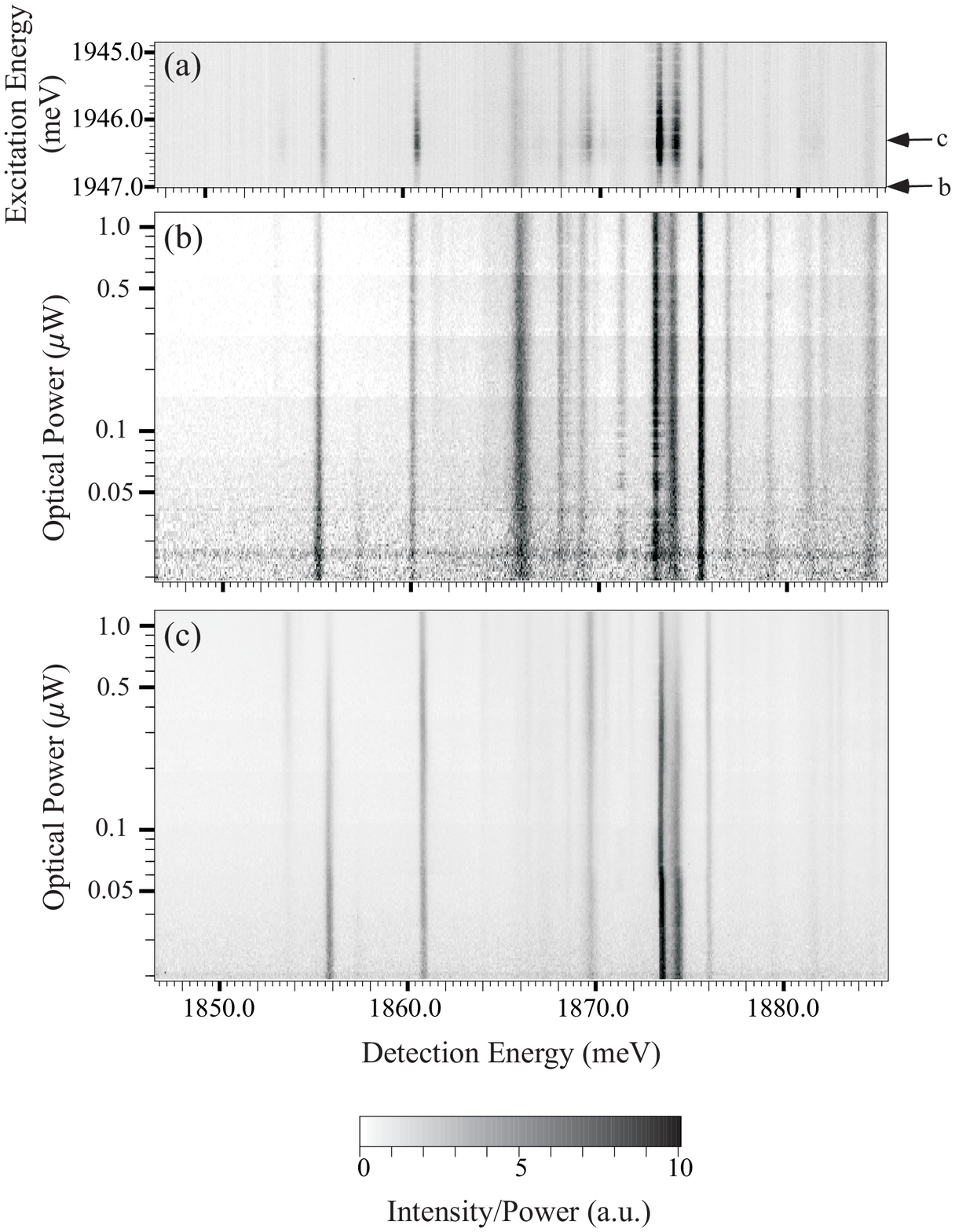,height=120ex}
\begin{center}
  Figure \ref{fig:powbWL}
\end{center}

\newpage
\psfig{figure=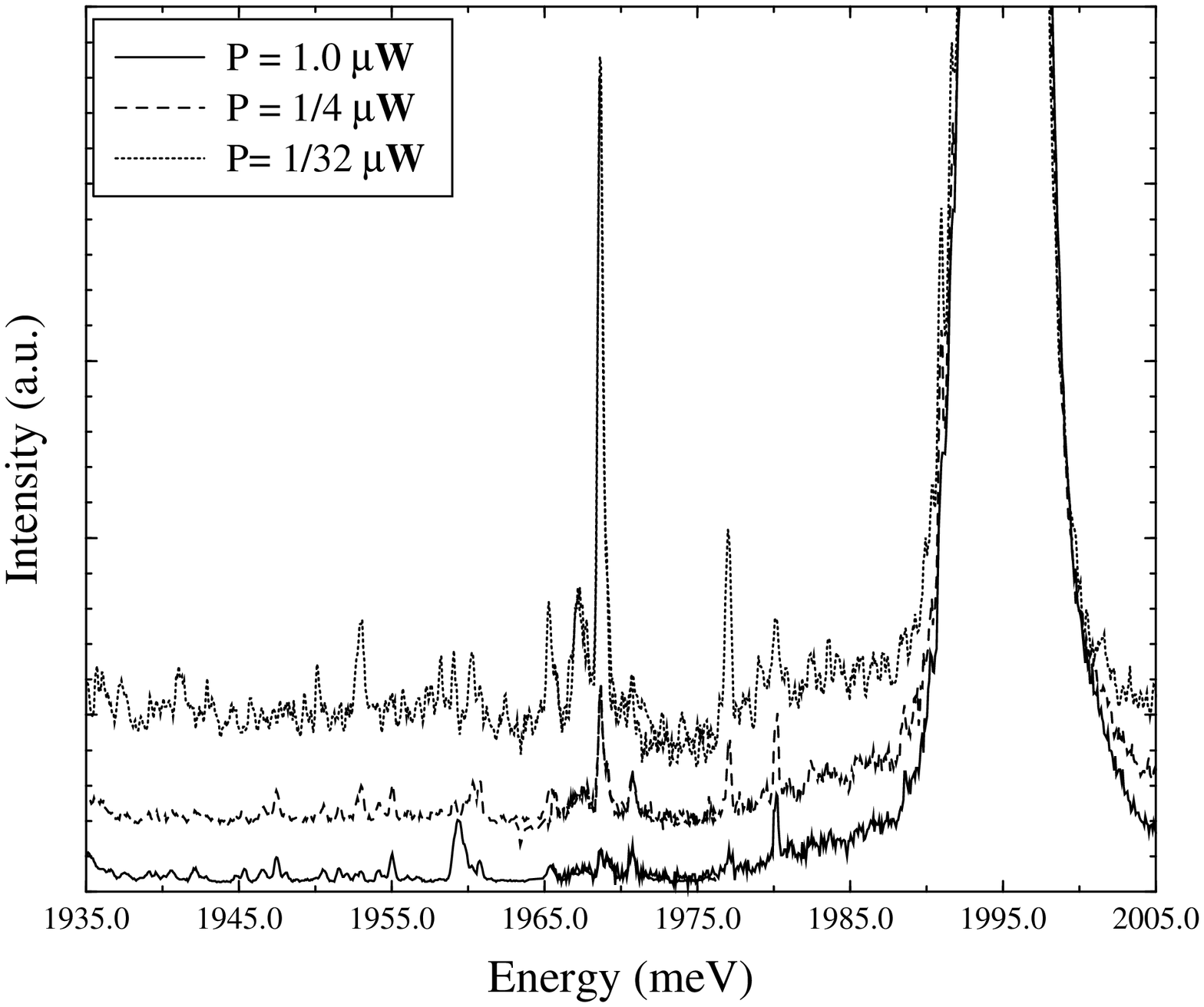,angle=-90}
\begin{center}
  Figure \ref{fig:roughstates}
\end{center}

\end{document}